\documentclass[%
 reprint,
superscriptaddress,
showpacs,showkeys,preprintnumbers,
 amsmath,amssymb,
 aps,
 prb,
]{revtex4-1}

\usepackage{graphicx}
\usepackage{dcolumn}
\usepackage{bm}
\usepackage[latin1]{inputenc}  
\usepackage{xcolor} 

\begin{document}

\title[Sample title]{Probing the local electronic structure of isovalent Bi atoms in InP}

\author{C. M. Krammel}
\affiliation{ 
Department of Applied Physics, Eindhoven University of Technology, Eindhoven 5612 AZ, The Netherlands
}%
\author{A. R. da Cruz}%
\affiliation{ 
Department of Applied Physics, Eindhoven University of Technology, Eindhoven 5612 AZ, The Netherlands
}%
\author{M. E. Flatt\'e}%
\affiliation{ 
Department of Physics and Astronomy, University of Iowa, Iowa City, Iowa 52242, United States of America
}%
\affiliation{ 
Department of Applied Physics, Eindhoven University of Technology, Eindhoven 5612 AZ, The Netherlands
}%
\affiliation{ 
Pritzker School of Molecular Engineering, University of Chicago, Chicago, Illinois, 60637, United States of America
}%

\author{M. Roy}
\affiliation{%
Department of Physics and Astronomy, University of Leicester, University Road, Leicester LE1 7RH, United Kingdom
}%
\author{P. A. Maksym}
\affiliation{%
Department of Physics and Astronomy, University of Leicester, University Road, Leicester LE1 7RH, United Kingdom
}%
\author{L. Y. Zhang} 
\affiliation{Department of Physics, University of Shanghai,for Science and Technology, Shanghai 200093, China}
\author{K. Wang} 
\author{Y. Y. Li}
\affiliation{State Key Laboratory of Functional Materials for Informatics, Shanghai Institute of Microsystem and Information Technology, Chinese Academy of Sciences, Shanghai 200050, China}

\author{S. M. Wang} 
\affiliation{State Key Laboratory of Functional Materials for Informatics, Shanghai Institute of Microsystem and Information Technology, Chinese Academy of Sciences, Shanghai 200050, China}
\affiliation{Department of Microtechnology and Nanoscience, Chalmers University of Technology, 41296 G\"oteborg, Sweden}

\author{P. M. Koenraad}%
\affiliation{ 
Department of Applied Physics, Eindhoven University of Technology, Eindhoven 5612 AZ, The Netherlands
}%

\date{\today}

\begin{abstract}
Cross-sectional scanning tunneling microscopy (X-STM) is used to experimentally study the influence of isovalent Bi atoms on the electronic structure of InP. We map the spatial pattern of the Bi impurity state, which originates from Bi atoms down to the sixth layer below the surface, in topographic, filled state X-STM images on the natural $\{110\}$ cleavage planes. The Bi impurity state has a highly anisotropic bowtie-like structure and extends over several lattice sites. These Bi-induced charge redistributions extend along the $\left\langle 110\right\rangle$ directions, which define the bowtie-like structures we observe. Local tight-binding calculations reproduce the experimentally observed spatial structure of the Bi impurity state. In addition, the influence of the Bi atoms on the electronic structure is investigated in scanning tunneling spectroscopy measurements. These measurements show that Bi induces a resonant state in the valence band, which shifts the band edge towards higher energies. Furthermore, we show that the energetic position of the Bi induced resonance and its influence on the onset of the valence band edge depend crucially on the position of the Bi atoms relative to the cleavage plane.
\end{abstract}

\maketitle

%

\section{Introduction}
\label{sec:Introduction}

Alloying conventional binary semiconductor materials with isovalent impurities, which are significantly different from the host in terms of size and electronegativity, provide a range of accessible perturbing features to the electronic structure without changing the local charge carrier density. In these materials, which are known as highly mismatched alloys (HMAs), the electronic structure of the parent material is perturbed by the impurity states leading to unusual electronic and optical properties. Prominent III-V compound based HMAs are dilute nitrides\cite{Francour/APL/72/1857/LuminescenceOfAsGrownAndThermallyAnnealedGaAsN,Shan/APL/76/3251/NatureOfTheFundamentalBandGapGaNPAlloys,Bi/Jap/80/NIncorporationinInPAndBandGapBowing} and dilute bismides\cite{Franceour/apl/82/3874/BandGapOfGaAsBi,Svensson/jvstb/30/MBEAndPLOfInAsBi,Wang/SciRep/4/5449/InPBSingleCrystalsGrownByMBE,Kopaczec/APL/ElectroreflectanceAndTheoreticalStudiesOfInPBi} which offer interesting prospects for new device concepts especially in the field of optical devices\cite{Tansu/APL/81/2523/Low-Treshold-Current-Density1300nmDilute-NitrideQuantumWellLasers,Sweeney/JAP/113/043110/Bismide-nitrideAlloys:PromisingForEfficientLightEmittingDevicesInTheNearAndMidInfrared,Wang/Springer/285/BIsmuthContainingAlloysAndNanostructures}. In these material systems, the band anticrossing (BAC) model is successfully used to describe in the dilute limit the influence of highly mismatched isovalent impurities on the band gap and the effective mass of HMAs. \cite{Wu/SemicondSciTechnol/17/869/BACInHighlyMismatchedIIIValloys,Shan/PhysRevLett/82/1221/BandAnticrossingInGaInNAsAlloys,Alberi/PhysRevB/75/045203/ValenceBandAnticrossingInMismatchedIIIVs} However, a better understanding of the electronic properties of HMAs beyond the phenomenological level provided by the BAC model requires detailed knowledge of the interactions of the isovalent impurities with each other as well as their host.  This has  primarily been investigated in a few fundamental works based on tight-binding (TB) and density functional theory (DFT) calculations, which involve real-space interpretations of the states introduced into the host by individual N and Bi atoms. \cite{Kent/PhysRevB/64/115208/TheoryOfElectronicStructureEvolutionInGaAsNAndGaPN,Virkkala/PhysRevB/88/035204/OriginOfBandGapBowingInDiluteGaAsNAlloys,Ivanove/PRB/82/161201/DirectMeasurementAndAnalysisOfTheCBDOSGaAsN,Virkkala/PhysRevB/88/235201/ModelingBiInducedChangesInTheElectronicStructure,Bannow/PhysRevB/93/205202/ConfigurationDependenceOfBandGapNarrowingAndLocalizationGaAsBi} 

Despite many parallels between dilute nitrides and dilute bismides, the nature of the states related to Bi and N dopants differs considerably.\cite{Zang/PhysRevB/71/155201/SimilarAndDissimilarAspectsOfIIIVSemiconductorsContainingBiAndN}  However, support from the experimental side for the single-added-atom structure remains minimal; most features seen experimentally emerge for an ensemble of added impurities. Recently cross-sectional scanning tunneling microscopy (X-STM) has been utilized to probe the electronic signatures of single N centers in GaAs.\cite{Ishida/Nanoscale/7/16773/VisualizationOfNInDiluteGaNAsByXSTM,Plantenga/PhysRevB/96/155210/SpatiallyResolvedElectronicStructureOfNitrogenCenterGaAs,Ivanove/PRB/82/161201/DirectMeasurementAndAnalysisOfTheCBDOSGaAsN}  N creates near the conduction band (CB) an $s$-like impurity state.\cite{Virkkala/PhysRevB/88/035204/OriginOfBandGapBowingInDiluteGaAsNAlloys,Virkkala/PhysRevB/88/235201/ModelingBiInducedChangesInTheElectronicStructure} About dilute III-V-bismides much less is known experimentally, although Bi is expected to introduce a mixed $p$-like state in the region of the valence band (VB).\cite{Virkkala/PhysRevB/88/235201/ModelingBiInducedChangesInTheElectronicStructure,Zang/PhysRevB/71/155201/SimilarAndDissimilarAspectsOfIIIVSemiconductorsContainingBiAndN} 
 In order to clarify the properties of single Bi atoms within III-V semiconductors, X-STM is used to investigate the electronic structure of the Bi impurity state near the $\left\{ 110 \right\}$ surfaces. 



In this paper, we probe the spatial structure of the Bi impurity state in cleaved $\{110\}$ InP surfaces. This is done in an approach where filled-state X-STM images for Bi atoms down to the sixth layer below the cleavage plane are analyzed.  The spatial structure of the experimentally observed Bi impurity state is compared to the spatial structure obtained in the TB calculations for Bi in InP. We find a much shorter-ranged perturbation of the valence band of InP relative to that calculated previously for Bi in GaAs.\cite{Virkkala/PhysRevB/88/235201/ModelingBiInducedChangesInTheElectronicStructure} 
We use scanning tunneling spectroscopy (STS) to investigate the local density of states (LDOS) of the Bi impurity states as a function of energy for Bi atoms at various depths below the surface.

\section{Experimental and computational methods}
\label{sec:Experimental and computational methods}
The investigated sample contains three different Bi-doped InP films (F) with Bi concentrations of $0.1~\%$ (F $1$), $0.5~\%$ (F $2$) and $1.0~\%$ (F $3$). Each composition is repeated three times in a superlattice-like InP:Bi(15 nm)/InP(20 nm) structure. Starting with the smallest Bi content, the three different sets of identical Bi doped films are separated by a  $100~\rm{nm}$ thick InP spacer layer. The superlattice region is terminated by a $50~\rm{nm}$ thick InP layer, which is followed  by a $150~\rm{nm}$ InP:Bi bulk layer with a Bi concentration of $2.4 \%$. This protects the region with the InP:Bi films (1, 2, 3) and allows easy orientation of the sample. Details of the growth of this samples can be found in an earlier work in which the same sample was investigated.\cite{Krammel/PRM/1/034606/IncorporationOfBiAtomsInPStudiedAtTheAtomicScaleByXSTM}

The X-STM measurements were performed in a commercial Omicron low temperature STM at 5 K. The samples were cleaved under ultra high vacuum (UHV) conditions in the STM chamber at pressures below $4\times 10^{-11}~\rm{mbar}$. Polycrystalline tungsten tips, which were electrochemically etched and further refined in the STM by Ar sputtering, were used as probes. All images were acquired in constant current mode at negative sample voltage. In this regime the tunneling current is dominated by the filled states in the VB, which are primarily associated with group V elements, such as P and Bi.\cite{Ebert/PhysRevLett/77/2997/SurfaceResonancesInSTM} 

To calculate the local density of states (LDOS) corresponding to the Bi atom we employed the Koster-Slater Green's function formalism \cite{kosterslaterPhysRev.95.1167} applied to a local defect\cite{tangflattePhysRevLett.92.047201,kortan2016} using a bulk $sp^{3}d^{5}s^{*}$ empirical tight-binding Hamiltonian\cite{jancuPhysRevB.57.6493} to describe the InP host. Results from this calculation method have  been compared previously to STM measurements of  single impurities  in III-V zinc-blende semiconductor hosts (e.g., Mn in GaAs \cite{Yakunin/PRL/92/216806/SpatialStructureOfIndividualMnAcceptorsInGaAs} and N in GaAs \cite{Plantenga/PhysRevB/96/155210/SpatiallyResolvedElectronicStructureOfNitrogenCenterGaAs}) showing good agreement. In our approach, the substitutional impurity potential is set by including on site energies equal to the difference in $s$, $p$, and $d$ atomic energy levels between bismuth and phosphorus, which are obtained from atomic spectra data.\cite{atomicspectra} The bond distance between the Bi and its four nearest-neighboring In atoms is set by tuning the hopping parameters according to  scaling laws described by Jancu. et al \cite{jancuPhysRevB.57.6493} by fitting strained bulk electronic structures.  This allowed us to place the bismuth resonance level at 120 meV below the valence band edge as estimated from previous calculations.\cite{tiwari1992empirical} The distance of the Bi-In bond is 35.3 \% longer than a bulk P-In bond.

\section{Results and Discussion}
\label{sec:Results and Discussion}

\subsection{Spatial structure of Bi impurity states }
\label{subsec:Spatial structure of Bi impurity states}

\begin{figure}[h]
\begin{center}                                       
\includegraphics[width=0.8\linewidth]{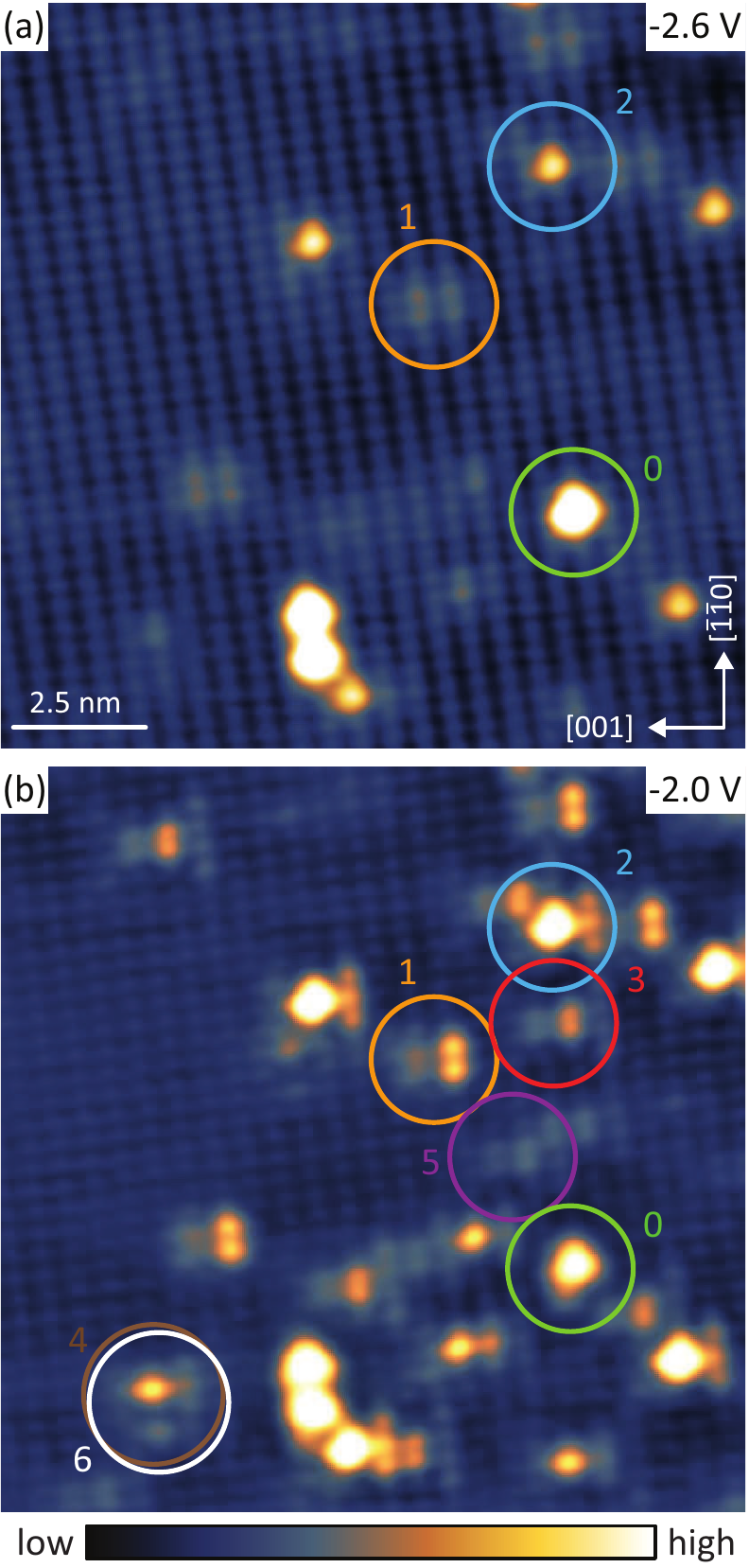} 
     \caption{Filled-state X-STM images of the same region of film $3$, which are acquired at a tunnel current set point of $I_T= 30~\rm{pA}$ on the $(\bar{1}10)$ InP surface. (a) X-STM image taken at a large negative sample voltage of $U_S=-2.6~\rm{V}$, where states deep in the VB are probed. In this case,  Bi atoms in the first three surface layers (0, 1, 2) are clearly visible. (b) X-STM image taken at a smalller negative sample voltage of $U_S=-2.0~\rm{V}$, where states closer to the VBE are probed. This results in a change of the appearance of the Bi atoms in layers (1) and (2). In addition, Bi atoms in deeper layers (3, 4, 5, 6) become visible. The color scales of both images (a) and (b) are independently adjusted for best visibility. The crystal directions are indicated in (a).}
     \label{fig:fig1}
\end{center}
\end{figure}

We use the ability of the STM to probe the LDOS at the atomic level to explore in conventional topographic filled state X-STM images the spatial structure of the Bi impurity state. This is complicated by the additional dependency of the tunnel current, $I_T$, on the real topography of the sample. In earlier work we have shown that the geometrical structure of the relaxed surface is considerably modified by Bi atoms down to the second layer below the cleavage plane.\cite{Krammel/PRM/1/034606/IncorporationOfBiAtomsInPStudiedAtTheAtomicScaleByXSTM,Tilley/PRB/ToBePublished} Therefore, it is important to make a clear distinction between topographic and electronic contributions from Bi atoms to the X-STM images. First indications on the role of these two effects can be derived from X-STM images of the same area, which are recorded at different sample voltages $U_S$.  Real topographic features affect the tunnel current irrespective of the applied voltage. This is due to the exponential dependence of the tunnel current on the distance. The LDOS of this sample has, in contrast to real topographic features, also a strong energy dependency which appears in the tunnel current only at specific sample voltages. In the case of isovalent Bi atoms in InP we have a resonant impurity state just below the valence band edge (VBE).\cite{Alberi/PhysRevB/75/045203/ValenceBandAnticrossingInMismatchedIIIVs} The Bi impurity state affects the tunnel current primarily at small negative sample voltages, when electrons are mainly extracted from occupied states near the VBE into empty tip states.  Therefore, it is sufficient to focus on the negative voltage range in which  $I_T$ is determined by the VB band states. 

Figure~\ref{fig:fig1} (a) shows a small part of  film 3, which is imaged at a large negative sample  voltage of $U_S=-2.6~\rm{V}$. Here, three different Bi related features can be distinguished (0, 1, 2), which arise from Bi atoms down to the second layer below the surface. The numbering starts with 0, which stands for the cleavage plane. The depth of the Bi atoms is determined on the basis of our earlier work in which we identify Bi atoms in and below cleaved $\{110\}$ InP surfaces.\cite{Krammel/PRM/1/034606/IncorporationOfBiAtomsInPStudiedAtTheAtomicScaleByXSTM} The same area of film 3 as in Fig.~\ref{fig:fig1} (a) is imaged in Fig.~\ref{fig:fig1}(b) at a reduced negative sample voltage of $U_S= -2.0~\rm{V}$. The  appearance of  Bi atoms in the first (1) and second (2) layer below the surface changes considerably at this tunneling condition. For example, two of the four P corrugation maxima, which are affected by a Bi atom in layer (1), are significantly brighter at small negative voltages than at larger voltages. Similarly, the atomic like contrast of Bi atoms in layer (2), which is identified at a large negative voltage, extends at smaller voltages over neighboring sites of the P corrugations. An exception are Bi atoms in the cleavage plane (0) whose contrast is hardly affected. In addition, previously unseen features, which are marked with numbered circles (3, 4, 5, 6), become visible. These features can only be found in Bi doped regions, which points towards impurity states of Bi atoms deep below the second layer ( e.g. 3, 4, 5, 6). The classification of these new Bi related features is explained in more detail later in this section. 

Bi atoms below layer (2) are hardly visible in Fig. \ref{fig:fig1} (a), which is taken at a large negative voltage where states deep in the VB far away from the Bi impurity state contribute most to the tunnel current. This illustrates that the geometric structure of the surface is primarily affected by Bi atoms down to the second layer (2) below the surface. Similarly, DFT calculations on the structural properties of Bi in InP show that the effectively larger Bi atoms give rise to a short ranged deformation extending only over a few lattice sites in the surrounding InP matrix.\cite{Tilley/PRB/ToBePublished,Krammel/PRM/1/034606/IncorporationOfBiAtomsInPStudiedAtTheAtomicScaleByXSTM} In contrast, the strong change in the appearance of the Bi atoms in layers (1) and (2) at a smaller negative sample voltage points to considerable contributions from the Bi impurity states in addition to the previously mentioned structural effects on the $\{110\}$ surfaces. Ultimately, the increased depth sensitivity, when addressing states close to the VBE, underlines that the contrast of the new Bi related features in layers 3, 4, 5, and 6 is almost entirely defined by the Bi defect state itself. An example of the sample bias-dependent evolution of the impurity state of a Bi atom in the third layer (3) is provided in section 1 of the Supplemental Material \cite{Supp}. Structural effects prevail for Bi atoms in the surface layer (0), whose appearance is not affected in the studied negative sample voltage range.

\begin{figure*}[htbp]
\begin{center}                                       
\includegraphics[width=1\linewidth]{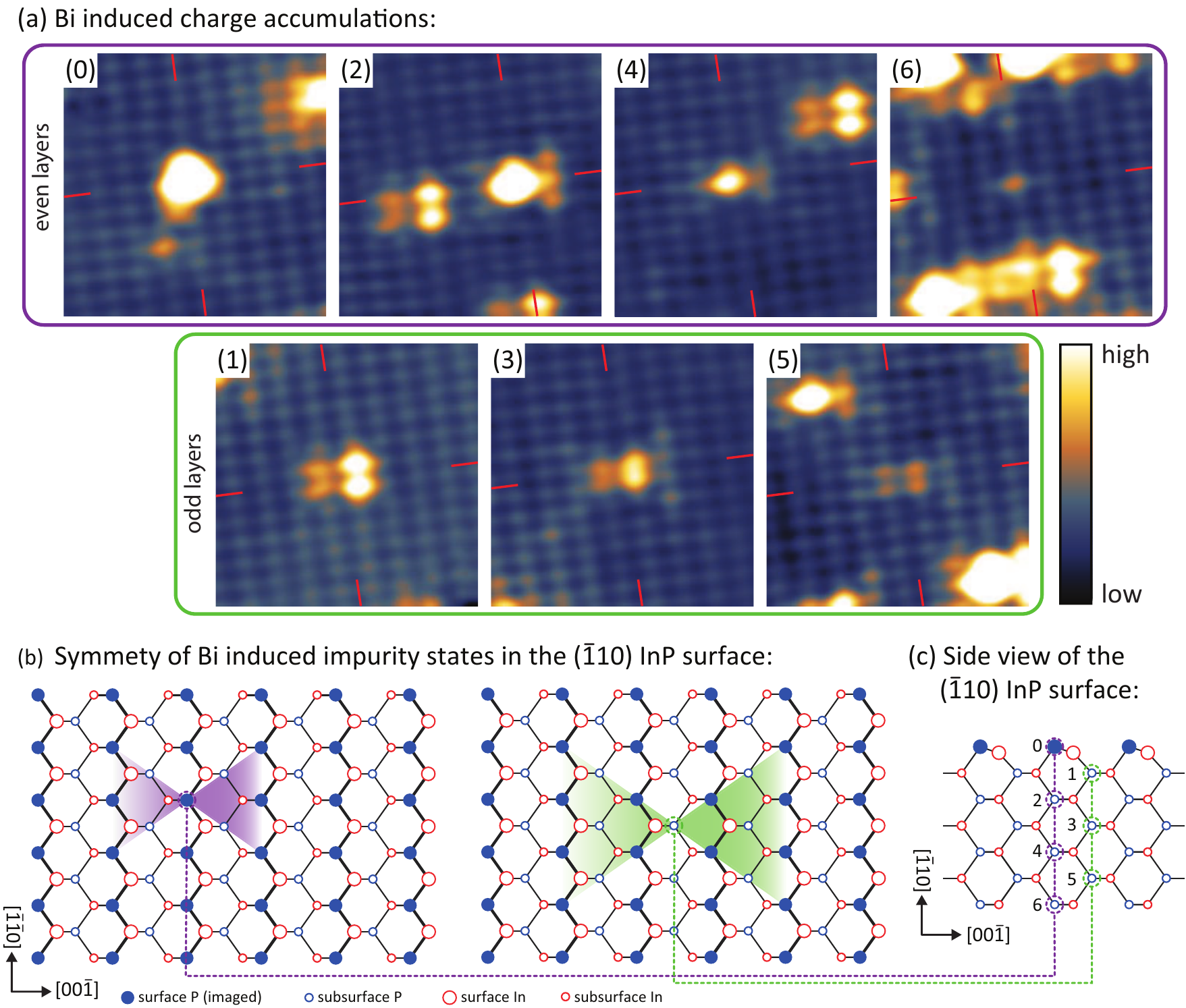} 
     \caption{  (a) High-resolution filled-state X-STM images of charge accumulations around individual Bi atoms in the first seven layers (0, 1, ..., 6) of the $(\bar{1}10)$ InP surface. The reference lines indicate the center of the Bi related features, which lie for Bi atoms in odd- or even-numbed layers on or in between the anion sublattice. All images are extracted from a large scale X-STM image of film $3$, which is taken at a tunnel current set point of $I_T= 40~\rm{pA}$ and a sample voltage of $U_S=-2.4~\rm{V}$. The color scales are independently adjusted to emphasize the spatial structure of the weaker features. (b) Schematic diagram showing the extent and symmetry of the charge accumulation around the Bi atoms in even and odd layers with respect to a top view of the $(\bar{1}10)$ InP surface. Here, filled blue disks represent the group V sublattice, which is probed in filled state X-STM images (a). (c) The positions of the Bi atoms in (a) are indicated in a side view of the relaxed $(\bar{1}10)$ surface.}
     \label{fig:fig2}
\end{center}
\end{figure*}

High-resolution filled-state X-STM images of the Bi impurity state, which originate from Bi atoms down to the sixth layer below the $(\bar{1}10)$ InP cleavage plane, are shown in Fig.~\ref{fig:fig2} (a). These maps taken under constant-current conditions resemble cuts along the cleavage plane through the three-dimensional Bi defect state, which are displaced by integer numbers of atomic layers (0, 1, 2, 3, 4, 5, 6) from the Bi atom. The Bi defect state displays, except for a Bi atom in the surface (0), a highly anisotropic structure, resembling the bowtie-like contrasts reported for deep acceptor wave functions in zinc-blende III-V compounds.\cite{Yakunin/PRL/92/216806/SpatialStructureOfIndividualMnAcceptorsInGaAs,Celse/PRB/77/075328/AnisotropicSpatialStructureOfDeepAcceptorStatesInGaAsAndGaP} However, acceptors such as Mn in GaAs, for which bowtie-like contrasts have been observed at an (110) cleavage plane, always induce a localized state in the band gap. In contrast to this, Bi in InP is an isovalent impurity, which gives rise to a resonant state outside the band gap just below the VBE.

The Bi impurity state is similar to deep acceptor states in that it is mirror symmetric in the $(\bar{1}\bar{1}0)$ plane. In the perpendicular directions, along the [001] directions, the observed symmetry of the Bi impurity states is broken. Height profiles, which cut through the highest point of the Bi impurity states in Fig.~\ref{fig:fig2} (a), are provided in Fig.~2 of the {supplementary} materials. The Bi impurity states can be divided into two types, which have their symmetry center either on or in-between the corrugation of the group V sublattice as indicated by reference lines in the first and second rows of Fig.~\ref{fig:fig2} (a). Sketches of these two configurations are provided in Fig. \ref{fig:fig2} (b) with respect to the underlying lattice. This points, in good agreement with our earlier work \cite{Krammel/PRM/1/034606/IncorporationOfBiAtomsInPStudiedAtTheAtomicScaleByXSTM}, to Bi atoms being incorporated at substitutional P sites in even- or odd-numbered layers as shown in Fig.~\ref{fig:fig2} (c). Bi is known to substitute for group V lattice positions in III-V compounds with zinc-blende structure, which are grown under group V rich conditions\cite{Wang/SciRep/4/5449/InPBSingleCrystalsGrownByMBE}. This allows us to attribute the new Bi related features based on the location of their symmetry center with respect to the group V sublattice in combination with their height in the STM image, which is expected to decrease with increasing distance from the surface, to Bi atoms on P sites between the third (3) and sixth (6) layers. Similarly, the maximum extension of the Bi impurity state increases along the [001] and $[\bar{1}\bar{1}0]$ directions in the $(\bar{1}10)$ surface by an additional lattice site when going from the surface (0) down to the third layer (3). The coupling of Bi atoms in odd-numbered layers to the surface is notably stronger than for Bi atoms in even numbered layers, as can be seen in Fig.~2 of the supplementary material. All this suggests that the strongly coupled zigzag rows of In and P atoms along the $\left\langle 110\right\rangle$ directions play an important role for the appearance of the Bi impurity states. A further increase of lateral extension of the impurity state is not observed beyond the third layer (3) below the cleavage surface.

\begin{figure*}[htbp]
\begin{center}                                       
\includegraphics[width=1\linewidth]{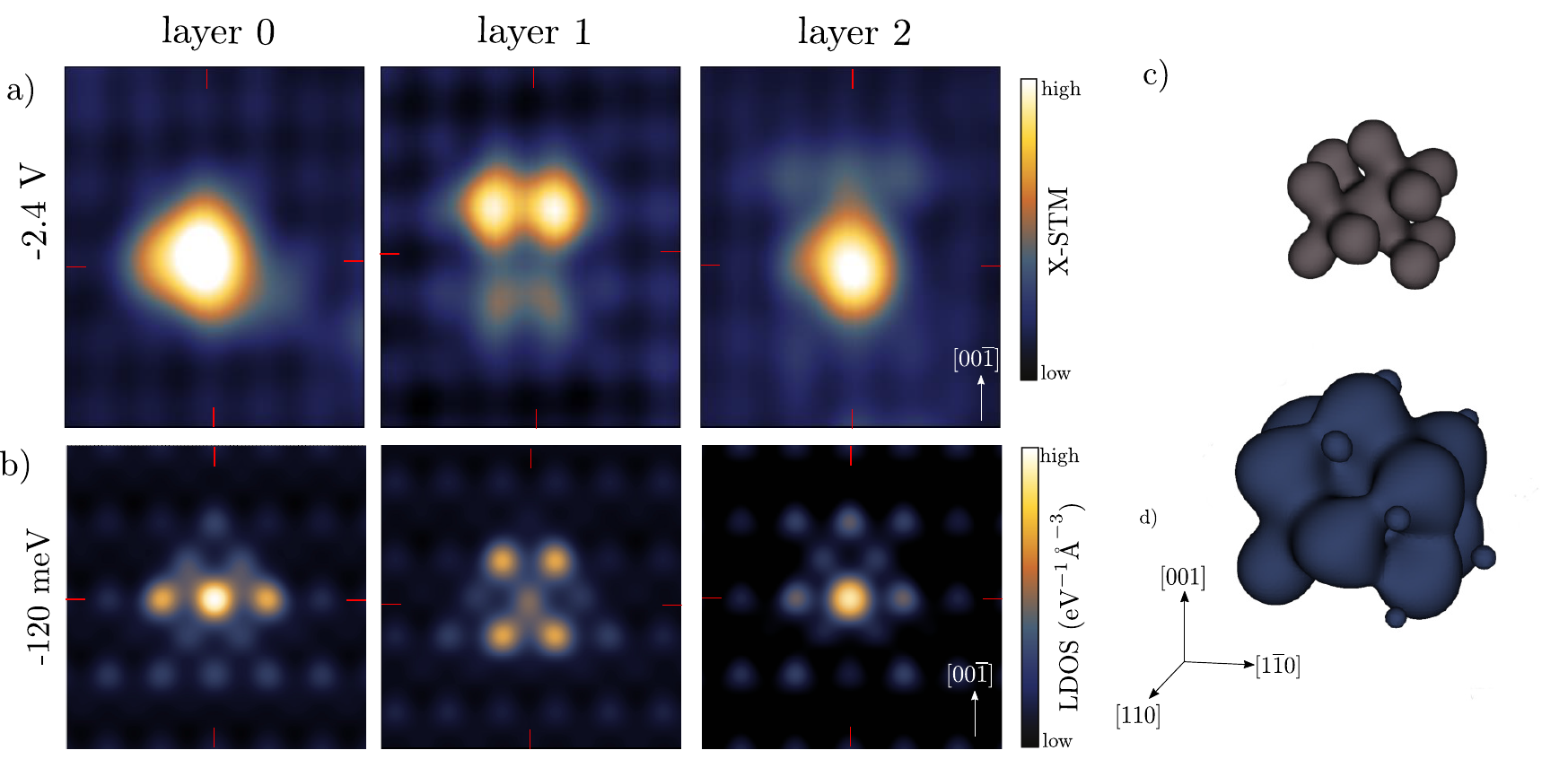} 
     \caption{(a) X-STM measurements of filled states of the $(\overline{1} 1 0)$ plane of InP taken at -2.4 V . The layers 0, 1 and 2 correspond to the cleavage surface and the two layers below it respectively. These correspond to the images (0), (1), and (2) of Fig.\ref{fig:fig3} (a). (b) LDOS calculated at 120 meV below the VBE with an assumed STM tip width of 1.13 \AA. The red reference lines indicate the location of the Bi atom. (c), (d) Isodensity surfaces calculated at 120 meV below VBE corresponding to a difference between the Bi atom and the background of 0.6 $\text{eV}^{-1}\text{\AA}^{-3}$ and 0.4 $\text{eV}^{-1}\text{\AA}^{-3}$ respectively. }
     \label{fig:fig6}
\end{center}
\end{figure*}

 In Fig.~\ref{fig:fig6} we compare the measured contrasts for filled-states X-STM images shown in Fig.~\ref{fig:fig6}(a) to our tight-binding calculations for the LDOS of Bi atoms, shown in Fig.~\ref{fig:fig6}(b).
Previous work has shown that discrepancies between X-STM contrasts and the theoretical calculated LDOS of defect states,
observed as a depth-dependent asymmetry of the wave function, can originate from strain induced by surface reconstruction which is stronger for the layers near the surface. \cite{Yakunin/PRL/92/216806/SpatialStructureOfIndividualMnAcceptorsInGaAs,Celse/PRB/77/075328/AnisotropicSpatialStructureOfDeepAcceptorStatesInGaAsAndGaP}. Since  our calculations were performed for a bulk system without this surface-induced strain, some asymmetries evident in the X-STM contrasts for layers close to the surface are not seen in the calculations. This discrepancy does not affect the other features of the calculations that, as we shall see below, are well reproduced by the calculations.

The calculations and measurements are shown in Fig.~\ref{fig:fig6} for the zeroth layer (0), corresponding to the $(\bar{1}10)$ cleavage plane, and the first (1) and second (2) layers below it. All calculated results were obtained for a resonant state at 120 meV below the VBE.  The spatial distribution of the LDOS at each atomic site was obtained by assuming the STM tip is well described by a Gaussian orbital whose width is 1.13 \AA, approximately half the distance between an In and its neighboring P atom. Features at smaller length scales could perhaps be fit better by evaluating the convolution between more accurate representations of tip orbitals and surface orbitals, but these features are not crucial to compare with X-STM image features on larger scales (which even includes atomic-scale properties). 

At the (110) surface of III-V semiconductors, a surface-relaxation-induced strain is responsible for the  breaking of the mirror symmetry for acceptors, {\it i.e.}, along the [001] direction for Mn dopant wave functions.\cite{Celse/PhysRevLett/104/086404/SurfaceInducedAsymmetryOfAcceptorWaveFunctions} Even in bulk there is a small symmetry-breaking effect in the (001) plane, but this is dwarfed by that induced by the surface strain. Thus, our calculations agree well with the measurements as long as the (001) reflection plane symmetry is (mostly) conserved, as we have not included surface strain in our calculations. The prominent deviations from  experiment in Figs.~\ref{fig:fig3}(a) and ~\ref{fig:fig3} (b) in layers 1 and 2 are consistent with the previously observed symmetry breaking, as the differences are characterized by a lower contrast symmetry along the [001] direction in the X-STM images than in the calculated LDOS.
In the simulation, which is performed for a bulk system, the enhancement of the LDOS by the Bi atoms is approximately symmetric, while in the experiment an asymmetry regarding the (001) plane is seen. The symmetry breaking observed in the experiment is thus due to the relaxation of the $(\bar{1}10)$ surface, which deforms the lattice and induces additional strain at the semiconductor-vacuum interface.\cite{Celse/PhysRevLett/104/086404/SurfaceInducedAsymmetryOfAcceptorWaveFunctions}

The global electronic properties of highly mismatched III-V-bismides, such as the band gap and spin orbit coupling, do not solely depend on the interactions between the Bi impurities and the host material. It has recently been suggested that in addition to perturbations induced by single Bi atoms also Bi-Bi interactions play  a crucial role in dilute bismides especially at higher Bi concentrations.\cite{Virkkala/PhysRevB/88/235201/ModelingBiInducedChangesInTheElectronicStructure,Bannow/PhysRevB/93/205202/ConfigurationDependenceOfBandGapNarrowingAndLocalizationGaAsBi} Figure \ref{fig:fig1} (b) shows that already in the range of a about $1~\%$ Bi the impurity states of a significant amount of Bi atoms are overlapping. The bowtie-like shape of the impurity state, which is observed for isolated Bi atoms, is also conserved for groups of Bi atoms nearby. This shows that the Bi-Bi interactions are highly anisotropic and the need to carefully address issues regarding coupled Bi atoms. The anisotropy of the Bi-Bi coupling in combination with the preferential formation of Bi pairs and clusters along the $\left\langle 110\right\rangle$ directions, which we report in Ref.\cite{Krammel/PRM/1/034606/IncorporationOfBiAtomsInPStudiedAtTheAtomicScaleByXSTM}, is expected to have a substantial influence on the strength of the Bi induced VB edge broadening compared to models, which are based on ideal crystals where the Bi atoms are distributed randomly.

\subsection{Bi induced modifications of the electronic structure}
\label{subsec:Bi induced modifications of the electronic structure}

\begin{figure}[htbp]
\begin{center}                                       
\includegraphics[width=1\linewidth]{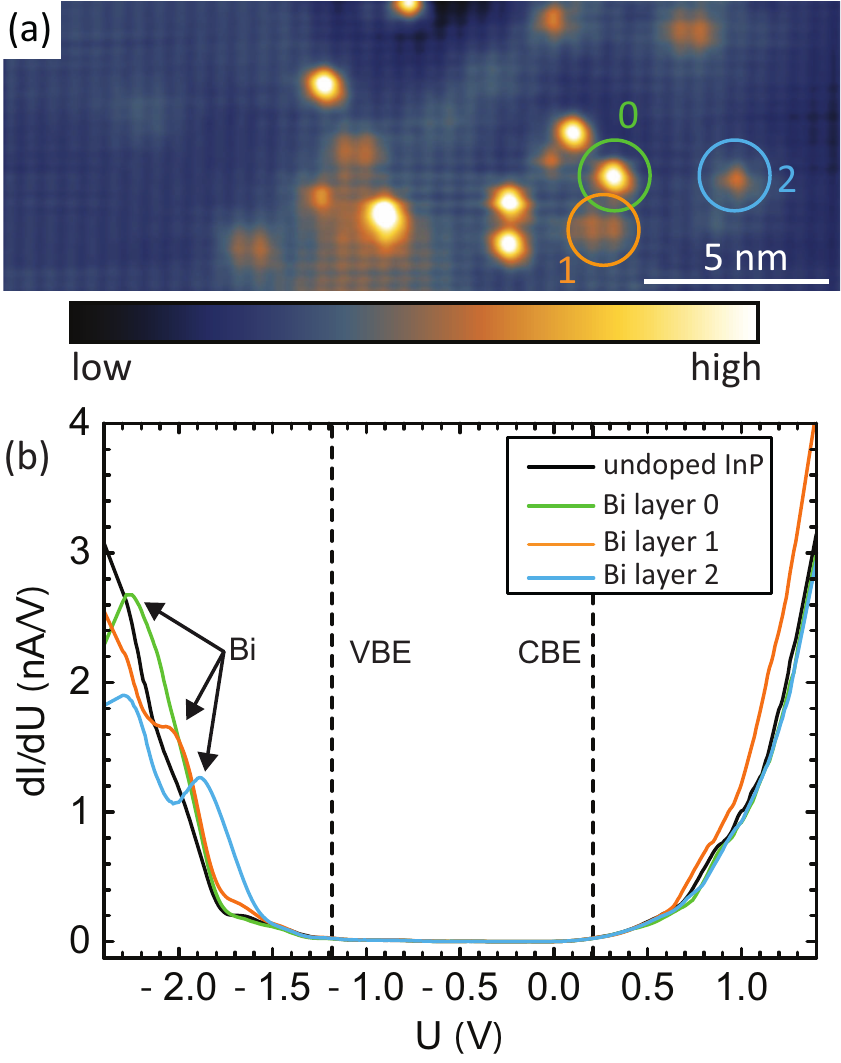} 
     \caption{ (a) Filled state X-STM image of film $3$ taken at a tunnel current set point of $I_T= 40~\rm{pA}$ and a sample voltage of $U_S=-2.4~\rm{V}$. (b) Spatially resolved differential conductance spectra [$dI(U)/dU$] taken at Bi atoms down to the second layer below the surface (0, 1, 2), which are marked in (a).  A spectrum acquired in the clean the InP matrix in between the Bi layers (black) serves as a reference. The spectra taken at Bi atoms show in contrast to the InP matrix additional resonances near the VBE, which lie deeper in the VB the closer the Bi atom is to the surface. The CB and VB edges are indicated by dashed vertical lines. The band gap is highlighted in gray. }
     \label{fig:fig3}
\end{center}
\end{figure}

\begin{figure}[htbp]
\begin{center}                                       
\includegraphics[width=1\linewidth]{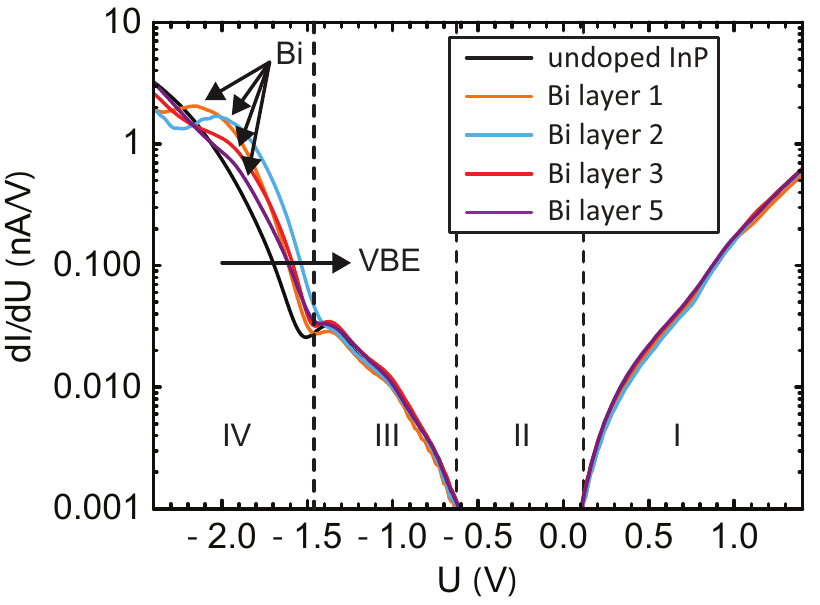} 
     \caption{The $dI(U)/dU$ spectra are extracted from a CITS map at the positions of Bi atoms in the first (1), second (2), third (3), and fifth (5) layer, which are labeled accordingly in 
     Fig. \ref{fig:fig1} (b). A spectrum of the bare InP matrix (black) serves as a reference. A logarithmic scale is used to reveal the VBE, which at the Bi atoms is shifted to higher energies as indicated by an arrow. The tunneling into the VB begins below -1.5 V (IV). The shoulder between -0.6 V and -1.5 V is related to a $I_T$ out of occupied CB states (III). In the band gap (II) tunneling is suppressed. Empty states in the conduction band are addressed at positive $U_S$ (I). The resonant states of Bi atoms in different layers are indicated by fanned arrows.}
     \label{fig:fig4}
\end{center}
\end{figure}

To further investigate the electronic properties of the Bi impurity state, differential conductance $dI(U)/dU$ spectra are taken at single Bi atoms down to the fifth  layer (5) below the surface. The  $dI(U)/dU$ can be seen as an approximation of the LDOS in the sample.\cite{Lang1986} We calculate the experimental $dI(U)/dU$ spectra on the basis of  $I(V)$ curves, which are afterwards numerically differentiated and smoothed with a Savitzky-Golay filter. The $I(V)$ curves are recorded with an open feedback loop after having approached the tip by an additional 0.16 nm towards the sample. Figure \ref{fig:fig3}  (b) shows typical $dI(U)/dU$ point spectra of individual Bi atoms down to the second layer below the surface and the surrounding InP matrix. These spectra are acquired in the back scan of the STM topograph in Fig. \ref{fig:fig3} (a), which is based on a tunnel set point of $I_T= 40~\rm{pA}$ and  $U_S=-2.4~\rm{V}$. The onsets of the CB and VB are indicated by dashed vertical lines, which are extracted from a logarithmic plot of the experimental $dI (U)/dU$ curves using a threshold of $t=0.026~\rm{nA/V}$.  Complementary $dI(U)/dU$ spectra of Bi atoms in deeper layers below the surface are provided in Fig. \ref{fig:fig4}. These $dI(U)/dU$ spectra represent averages of the positions of the correspondingly labeled Bi atoms in Fig. \ref{fig:fig1} (b), which are extracted from an $I(U)$ map. In order to reduce the topographic crosstalk in the spectroscopic curves shown in Fig. \ref{fig:fig4} we used the tunnel set point of $I_T= 30~\rm{pA}$ and $U_S=-2.5~\rm{V}$. Under these tunnel conditions Bi atoms below the second layer are hardly visible and the topographic contrast on Bi atoms in the surface layer and one layer below the surface is minimized, which limits topographical interference as much as possible in this material system. 

The logarithmic scale in Fig. \ref{fig:fig4}, which increases the weight of the weaker data points, shows that four tunneling regimes can be distinguished.\cite{Ishida/PRB/80/075320/InfluenceOfSurfaceStatesOnTunnelingSpectra} They are characteristic for the investigated sample. Tunneling from filled tip states into empty CB states takes place at positive $U_S$ in region (I). The fact that the tunneling in Region I starts at 0~V is consistent with the $n$-type character of the grown material which is common for low temperature grown InP and Bi:InP\cite{Gelczuc/PRB/49/115107/BiInducedAcceptorLevel}. Region (II) reflects the band gap, where no states are available. The shoulder in region (III) originates from an electron current out of a tip induced charge accumulation layer in the CB. This is related to tip induced band bending (TIBB), which pulls  the conduction band edge (CBE) below the Fermi level of the sample at a large enough negative $U_S$. Estimation of the TIBB for our n-type InP material shows an upward band bending of about 200 meV at $U_S=0~\rm{V}$  whereas the flat band condition is estimated to occur at about $U_S=-1~\rm{V}$ exactly where the surface accumulation will start that is responsible for the tunnel current in region (III). The additional increase in the $dI(U)/dU$ spectra at higher negative $U_S$ in region (IV) is due to contributions from electrons of the VB to the $I_T$. 

\begin{figure}[htbp]
\begin{center}                                       
\includegraphics[width=1\linewidth]{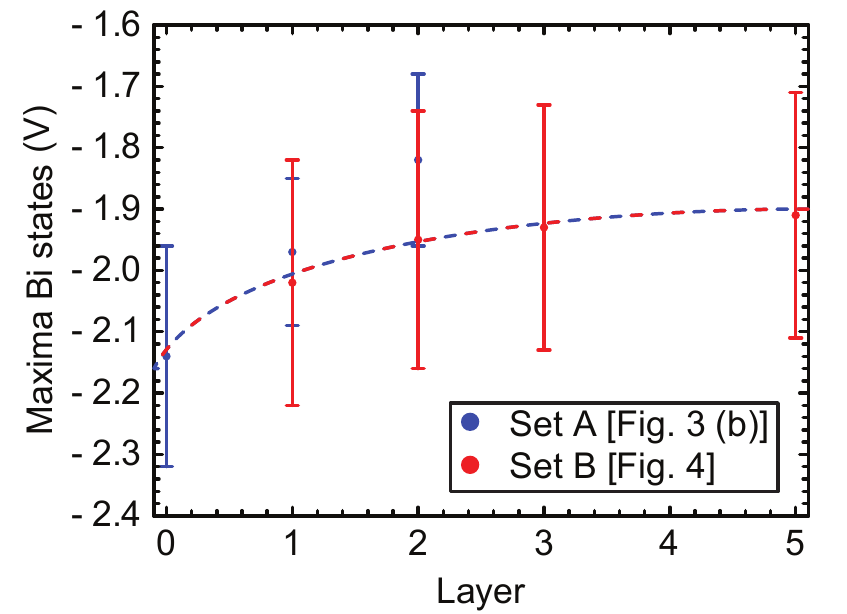} 
     \caption{ Depth dependency of the energetic positions of Bi impurity state in the VB. Set A, which is shown in blue, is extracted from the  $dI(U)/dU$ spectra in Fig. \ref{fig:fig3} (b). Set B, which is shown in red, is determined on the basis of the $dI(U)/dU$ spectra in Fig. \ref{fig:fig4}. The energetic positions of the Bi-related resonances in the experiment are extracted after subtraction of the InP background. The error bars represent the FWHM of the Bi impurity states.}
     \label{fig:fig5}
\end{center}
\end{figure}

The spectra taken at Bi atoms in different layers and the InP matrix agree very well in regions I, II, and III, which are dominated by the LDOS in the CB. This excludes a strong interaction of the Bi atoms with the CB states. Conversely, the electronic structure of the VB, which is represented by region IV, is heavily affected by the presence of Bi. Here, specific Bi-related resonances are observed, which are not present in the $dI(U)/dU$ spectra of the InP matrix. This reflects the acceptor like nature of the Bi impurity state. However, acceptor states lie typically a few tens of meV, depending on the species and host material, above the VBE in the band gap of III-V semiconductors. The magnitude of the Bi defect states can be judged best with the linear scale used in Fig. \ref{fig:fig3} (b). Interestingly, the Bi defect states, which originate from Bi atoms at different depths below the surface, do not lie equally deep in the VB as expected for single Bi atoms far away from each other and any interfaces. The depth-dependence of the energetic positions of the Bi impurity states in the experimental $dI(U)/dU$ spectra of Figs. \ref{fig:fig3} (b) and \ref{fig:fig4} is summarized in Fig. \ref{fig:fig5}. The energetic positions of the Bi impurity states are determined after having subtracted the InP background. In this way we compensate for energy shifts by the superimposed VB states. The error bars represent the full width at half maximum (FWHM) of the Bi states after the background subtraction. The Bi defect states in layers (0, 1, 2) shift with increasing distance from the surface closer to the VBE. This trend is not continued for Bi atoms below the second layer (2) where a stabilization of the Bi impurity state occurs at about $(1.9 \pm 0.1)~\rm{V}$. Interestingly an additional resonance seems to appear in our spectroscopic window of analysis for impurities at least two monolayers below the 110 surface. Unfortunately the tip and sample stability did not allow to pursue these resonances at more negative voltages.   

It is possible to exclude that the experimentally observed Bi defect states, which are related to Bi atoms in the first three layers (0, 1, 2), are shifted due to changes of the tip state, interactions between Bi atoms close by, or local potential fluctuations in the surrounding of the Bi atoms. This follows from the observation of corresponding spectral features with different W-tips at Bi atoms, which are further apart from each other than the ones discussed here (see supplementary material for more spectroscopic curves obtained on different Bi atoms). In addition, the topography of Bi atoms in the layers (0, 1, 2), which prevails at high negative $U_S$, leads inevitably to variations in the tip height. This can produce topographic cross talk in the STS spectra due to the additional dependence of the differential conductance on the tip-sample distance. A larger/smaller tip sample distance leads to an underestimate/overestimate of the local density of states (LDOS) and thus to an effectively lower/higher differential conductance.\cite{Wijnheijmer/10/4874/NanoscalePotentiaFluctuationsIn(GaMn)As-GaAs} The experimental $dI(U)/dU$ spectra of the InP matrix and Bi atoms agree well at positive voltages, which excludes a strong topographic cross talk. However, when viewing the $U$-scale of the experimental $dI/dU(U)$ spectra in absolute terms, one has to be cautious. The potential applied between the sample and tip, $U_S$, partially extends into the semiconductor, which gives rise to tip-induced band bending (TIBB).\cite{Feenstra/Nanotechnology/18/440156/InfluenceOfTIBBOnSTSSpectra} The TIBB acts as an additional lever, which affects the energy scale in the experimental $dI(U)/dU$ spectra. However if states shift due the fact that the TIBB is modified by the increased tip distance on top of a Bi atom (as seen in the topography image) this would result in a smaller TIBB and thus the resonance in the spectroscopic I(V) curve would appear at a lower voltage. This is opposite to what we observe where the resonances shift to larger negative voltages for Bi atoms closer to the surface. Therefore we can conclude that the modified lever-arm due to topographic crosstalk is not responsible for the observed shift of the resonance for Bi atoms at different depths below the cleavage plane. 

It seems likely that the energy shift of the Bi impurity state in layers 0, 1, and 2 is related to the semiconductor-vacuum interface itself. For instance, atoms in the surface have only three bonds to neighboring atoms. The fourth bond is a dangling bond, which is created by the cleavage. This is a completely different situation than in the crystal. In addition, all $\{110\}$ surfaces of III-V semiconductors with a zinc-blende structure relax with the anions moving outwards and the cations moving inwards, which also influences the arrangement of the atoms in the next lower layer.\cite{Chelikowsky/PRB/20/4150/SelfConsistentPseudopotentialCalculationsForTheRelaxedSurfaceOfGaAs} These structural modifications come along with electronic changes. Similarly, Bi atoms in deeper layers are more constrained by the surrounding lattice, which is increasingly difficult to deform. This leads to growing compressive strain at Bi atoms deeper in the crystal. Under pressure the Bi state is typically shifted deeper into the VB.\cite{Zang/PhysRevB/71/155201/SimilarAndDissimilarAspectsOfIIIVSemiconductorsContainingBiAndN}  Another relevant effect is that, the effective dielectric constant of the semiconductor near the surface is expected to be lower than far away from the surface. These effects are strongest near the surface, which fits well to the energetic shift of the Bi impurity states in the first three layers compared to the ones in deeper layers. A similar deepening of the impurity state close to the surface has been reported before\cite{Wijnheijmer2009,Strandberg2010,Garleff2010}.

Close inspection of the $dI(U)/dU$ curves in Fig. \ref{fig:fig4} reveals that Bi not only gives rise to a resonance in the VB but also reduces the band gap locally. This is caused by a shift of the VBE in region (IV) to higher energies, which is reflected in an earlier onset of region (IV). A change of the structure of the VB in regions (I) and (III) is not observed. Interestingly, the magnitude of the VB edge extension into the gap varies for Bi atoms at different depth, which we attribute to two effects. On the one hand, the perturbation of the host states decays with growing distance from the Bi atoms. On the other hand, we have shown that the Bi impurity state near the surface is shifted deeper into the VB, which in turn reduces its effect on the VBE. These two opposing effects cause, that the VB edge extension is strongest for a Bi atom in the second layer (2) below the surface. The VBE at the Bi atom in layer (2) is between -1.52 V and -1.85 V on average shifted by $\Delta_{VBE}=(0.21\pm0.03)~\rm{V}$ towards higher energies, which has to be seen as a lower limit for the Bi induced lifting of the VBE. The modification of the VBE is expected to be largest directly at the Bi atom, which corresponds to Bi atoms in the surface (1). But this cannot be measured directly due to the additional influence of the surface on the position of the Bi defect state.

Many band models, which predict the influence of Bi on the band gap of III-V semiconductor compounds, suggest that at low Bi concentrations there is a homogeneous shift of the VBE towards higher energies.\cite{Alberi/PhysRevB/75/045203/ValenceBandAnticrossingInMismatchedIIIVs,Batool/JAP/111/113108/ElectronicBandStructureOfGaBiAsAndGaAsLayers,Polak/SemicondSciTechnol/30/094001/FirstPrinciplesCalculationsOfBismuthInducedChangesInTheBandStructure,Usman/PRB/84245202/TBAnalysisOfTheElectronicStructureOfBismideGaAsAndGaP} Our experiment shows the nature of the band gap reduction in these regimes is rather determined by local Bi induced potential fluctuations, which give rise to localization. In contrast to some theoretical work,\cite{Batool/JAP/111/113108/ElectronicBandStructureOfGaBiAsAndGaAsLayers,Polak/SemicondSciTechnol/30/094001/FirstPrinciplesCalculationsOfBismuthInducedChangesInTheBandStructure,Usman/PRB/84245202/TBAnalysisOfTheElectronicStructureOfBismideGaAsAndGaP} which suggest that Bi incorporation affects both the CB and VB, we see at the atomic level no indications for a downward shift of the VB by single Bi atoms. The observation of a Bi-related resonance in the VB in combination with the VB edge extension is more in favor of a conventional BAC interaction.

\section{Conclusion}
\label{sec:Conclusion}

In this work, we close the gap between theory, which discusses the perturbation of conventional III-V based semiconductor alloys by isovalent Bi impurities, and experimental observations, which are largely lacking in real-space imaging at the atomic level. We address this by utilizing X-STM to investigate the influence of Bi impurities on the charge-carrier distribution of InP.
We visualize in filled-state X-STM images of the natural $\{110\}$ cleavage plane the depth-dependence of the Bi impurity state.  The depth of the Bi actual atoms is determined on the basis of the symmetry, height, and extensions of the Bi impurity states. Voltage-dependent measurements show that topographic modifications of the surface by the effectively larger Bi atoms are only relevant for Bi atoms down to the second layer below the surface. In fact, the Bi impurity state has a highly anisotropic bowtie-like shape and extends over several lattice sites, which points to a strong coupling with the host along to the zigzag rows of In and P atoms in the $\left\langle 110\right\rangle$ directions. The structure of the Bi resonant state is well reproduced by tight-binding calculations incorporating the atomic shifts of levels at the Bi impurity, along with a small shift of the Bi-In bond length relative to the bulk P-In bond length. Small discrepancies can be understood as due to the surface strain induced symmetry breaking of the valence state electronic structure.

Complementary STS measures on Bi atoms at different depths below the surface reveal distinct Bi-related resonances in the VB and a shift of the band edge towards higher energies.  In the first layers (0, 1, 2) near the surface the energetic position of the Bi impurity state is increasingly affected by the semiconductor-vacuum interface. This leads to a progressive shift of the Bi impurity state further into the valence band with increasing proximity to the surface. As a result, the displacement of the VBE with $\Delta_{VBE}=(0.21\pm0.03)~\rm{V}$ is most pronounced for Bi atoms in the second layer (2) below the surface, which has to be seen as a lower limit for the Bi induced lifting of the VBE.

\section*{Acknowledgments}

CMK and PMK thank NanoNextNL, a microtechnology and nanotechnology consortium of the Government of the Netherlands and $130$ partners for financial support. LYZ, KW, YYL and SMW wish to acknowledge the National Basic Research Program of China (Grant No. $2014$CB$643902$) and the Key Program of Natural Science Foundation of China (Grant No. $61334004$) for financial support. This project has received funding from the European Union's Horizon 2020 research and innovation programme under the Marie Sklodowska-Curie Grant Agreement No 721394. Contributions to the development of the theoretical calculation by M.E.F. for Bi in InP through modification of hopping matrix elements associated with changes in bond length were supported by the  U.S. Department of Energy, Office of Basic Energy Sciences, Division of Materials Sciences and Engineering under Award No. DE-SC0016447 and its renewal Awarad No. DE-SC0016379.

\nocite{*}
 \newcommand{\noop}[1]{}

\end{document}